\documentclass[10pt,conference]{IEEEtran}
\IEEEoverridecommandlockouts

\def\BibTeX{{\rm B\kern-.05em{\sc i\kern-.025em b}\kern-.08em
    T\kern-.1667em\lower.7ex\hbox{E}\kern-.125emX}}

\makeatletter
\newcommand{\linebreakand}{%
  \end{@IEEEauthorhalign}
  \hfill\mbox{}\par
  \mbox{}\hfill\begin{@IEEEauthorhalign}
}
\makeatother

\usepackage{cite}

\usepackage{amsmath,amssymb,amsfonts}
\usepackage{mathtools}

\usepackage{minted}
\usepackage{xspace}
\usepackage{float}
\usepackage{listings}
\usepackage{color}
\usepackage[table]{xcolor}      
\usepackage{pdfpages}
\usepackage{graphicx}
\usepackage{wasysym}
\usepackage{caption}
\usepackage{subcaption}
\usepackage[ruled,vlined,linesnumbered]{algorithm2e}
\usepackage{enumitem}
\usepackage{multirow}
\usepackage{booktabs}           
\usepackage[misc]{ifsym}
\usepackage[many]{tcolorbox}
\tcbset{
  boxsep=1pt,
  left=4pt,
  right=4pt,
  top=2pt,
  bottom=2pt,
  fontupper=\small,
  before skip=3pt,
  after skip=3pt,
}
\usepackage{balance}

\usepackage{hyperref}

\newcommand{\distance}{5pt}
\newcommand*{\tool}{{\textsc{Knowdit}}\xspace}

\newcommand{\chong}[1]{{#1}\xspace}
\newcommand{\ziqiao}[1]{{#1}\xspace}
\newcommand{\wanxu}[1]{{#1}\xspace}

\begin{document}

\title{\tool: Agentic Smart Contract Vulnerability Detection with Auditing Knowledge Summarization}

\author{%
\IEEEauthorblockN{Ziqiao Kong}
\IEEEauthorblockA{\textit{Nanyang Technological University} \\
Singapore, Singapore \\
ziqiao001@e.ntu.edu.sg}
\and
\IEEEauthorblockN{Wanxu Xia}
\IEEEauthorblockA{\textit{Beihang University} \\
Beijing, China \\
ysiel@buaa.edu.cn}
\and
\IEEEauthorblockN{Chong Wang\thanks{Chong Wang is the corresponding author.}}
\IEEEauthorblockA{\textit{Nanyang Technological University} \\
Singapore, Singapore \\
chong.wang@ntu.edu.sg}
\linebreakand
\IEEEauthorblockN{Yue Xue}
\IEEEauthorblockA{\textit{Independent Researcher} \\
Singapore, Singapore \\
nerbonic@gmail.com}
\and
\IEEEauthorblockN{Yi Lu}
\IEEEauthorblockA{\textit{Movebit} \\
Singapore \\
y@movebit.xyz}
\and
\IEEEauthorblockN{Pan Li}
\IEEEauthorblockA{\textit{Bitslab} \\
Singapore \\
paul@bitslab.xyz}
\linebreakand
\IEEEauthorblockN{Shaohua Li}
\IEEEauthorblockA{\textit{The Chinese University of Hong Kong} \\
Hong Kong, China \\
shaohuali@cuhk.edu.hk}
\and
\IEEEauthorblockN{Cao Zong}
\IEEEauthorblockA{\textit{Imperial Global Singapore} \\
Singapore \\
z.cao@imperial.ac.uk}
\and
\IEEEauthorblockN{Yang Liu}
\IEEEauthorblockA{\textit{Nanyang Technological University} \\
Singapore, Singapore \\
yangliu@ntu.edu.sg}
}

\maketitle

\begin{abstract}
Smart contracts govern billions of dollars in decentralized finance (DeFi), yet automated vulnerability detection remains challenging because many vulnerabilities are tightly coupled with project-specific business logic.
We observe that recurring vulnerabilities across diverse DeFi business models often share the same underlying economic mechanisms, which we term \textit{DeFi semantics}, and that capturing these shared abstractions can enable more systematic auditing.
Building on this insight, we propose \tool, a knowledge-driven, agentic workflow for smart contract vulnerability detection. \tool first constructs an auditing knowledge graph from historical human audit reports, linking fine-grained DeFi semantics with recurring vulnerability patterns. Given a new project, a multi-agent pipeline leverages this knowledge through an iterative loop of specification generation, \ziqiao{Proof-of-Concept (PoC) synthesis, PoC execution,} and finding reflection, driven by \ziqiao{a shared repository index}.

We evaluate \tool on \ziqiao{11 recent Code4rena projects with 84 ground-truth vulnerabilities. \tool detects all 21 high-severity and 90\% of medium-severity vulnerabilities without false positives, fully covering 8 projects, significantly outperforming all baselines. Applied to seven real-world projects, \tool further discovers 9 high- and 36 medium-severity previously unknown vulnerabilities, securing millions in liquidity and proving its outstanding performance.}
\end{abstract}

\begin{IEEEkeywords}
smart contract security, vulnerability detection, large language model agent, knowledge graph
\end{IEEEkeywords}

\section{Introduction}

Smart contracts are stateful programs deployed on blockchains that manage assets worth billions of dollars.
Due to the transparency and immutability of blockchain systems, vulnerabilities in smart contracts can lead to significant financial losses~\cite{zhou2023sok}.
For example, a recent security report\footnote{https://www.certik.com/blog/hack3d-the-web3-security-report-2025} by CertiK shows that over \$3.35 billion in cryptocurrency was stolen through hacks and scams in 2025 alone.
To mitigate these risks, automated analysis techniques, such as static auditing~\cite{gptscan} and fuzz testing~\cite{smartian,ityfuzz,verite}, have been proposed to detect bugs and security vulnerabilities in smart contracts and thereby safeguard on-chain assets.

Recently, large language models (LLMs) have shown promising results across a range of code intelligence and security tasks, including smart contract vulnerability detection~\cite{evmbench}.
GPTScan~\cite{gptscan} pioneers in matching smart contract code with potential attacking scenarios to reason about potential vulnerabilities.
Several works~\cite{gptscan,promfuzz,combingfinetuning,smartinv} leverage LLMs' ability to understand business models and generate invariants to detect smart contract vulnerabilities.
SmartPoc~\cite{smartpoc} and \texttt{A1}~\cite{a1} rely on LLMs for detailed reasoning and execution to synthesize exploits for smart contract vulnerabilities.

Despite these advances, automated smart contract auditing is still rather challenging.
A key reason is that many vulnerabilities are tightly coupled with specific business scenarios~\cite{web3bugs,promfuzz}, making it difficult to systematically model DeFi security properties and leverage them for comprehensive auditing processes and effective detection oracles.
Unlike traditional software vulnerabilities (e.g., \textit{use-after-free}), smart contract vulnerabilities often emerge within DeFi protocols that implement highly customized domain-specific business logic~\cite{web3bugs}.
This logic governs asset flows, token accounting, and complex financial operations.
As a result, the vast space of possible DeFi scenarios and auditing objectives creates significant challenges for automated detection, particularly due to the gap between accurately capturing contract semantics and applying appropriate auditing workflows and oracles.
For example, generic fuzzing oracles like \textit{profit generation} or \textit{reentrancy} are often insufficient to cover the wide variety of smart contract auditing scenarios.

Our key insight is that the high false-negative rate largely stems from the lack of high-level abstractions that capture the shared auditing knowledge underlying diverse DeFi protocols and their potential vulnerabilities.
We define the fine-grained functionalities that implement specific economic mechanisms as \textit{\textbf{DeFi semantics}}~\cite{defiranger}.
In practice, many vulnerabilities originate from common DeFi semantics embedded in protocol business logic, along with the attack scenarios that exploit them. While protocol designs and concrete implementations may differ due to varying business requirements, the underlying economic mechanisms represented by these DeFi semantics—and their associated attack surfaces—often recur across protocols.
For example, two smart contract vulnerabilities discovered in \textit{InsureDAO} in 2022 and \textit{Salty.IO} in 2024 illustrate this recurring pattern \ziqiao{as our Section~\ref{section:motivation} suggests}.
Although these vulnerabilities appear in entirely different protocols—one focused on insurance underwriting and the other on automated market-making and liquidity management—they both stem from the same DeFi semantic pattern, \textit{proportional-share token accounting}, and share a common attack scenario known as the \textit{first-depositor attack}.
This observation underscores the importance of leveraging high-quality, abstract auditing knowledge that captures both shared DeFi semantics and the attack scenarios they enable. By modeling these essential properties, rather than being distracted by superficial differences in implementation, we can enable more effective vulnerability detection.

However, effectively utilizing such auditing knowledge presents several challenges.
First, there is a lack of systematic understanding and modeling of the essential knowledge elements and the relationships among them during smart contract auditing.
Second, current automated methods are inadequate for fully extracting and evolving the auditing knowledge from heterogeneous sources.
Many existing works~\cite{gptscan,promfuzz,propertygpt,smartinv} either rely on the fixed template or extra manual efforts for such knowledge, preventing the large-scale adoption of the workflows.
Third, accurately mapping abstract auditing knowledge to concrete smart contract implementations and performing vulnerability detection remains challenging for general and scalable automated auditing~\cite{defiranger,promfuzz}.

To address these challenges, we propose a novel knowledge-driven, agentic method for smart contract vulnerability detection, called \tool.
Our method consists of two main phases. First, we construct an auditing knowledge graph $\mathcal{G}$ from historical human reports based on a schema that separates the DeFi Space, which captures business types, Solidity projects, and fine-grained DeFi semantics, and the Vulnerability Space, which encodes vulnerability patterns, auditing findings, and attack types, with links representing historical traceability and potential causal relationships.
The graph is built incrementally using an LLM-based pipeline that abstracts, classifies, and deduplicates knowledge.
Second, given a new Solidity project, our agentic auditing framework uses a shared Repository Index to track execution, coverage, and feedback.
It maps the project to relevant DeFi semantics and vulnerability patterns and iteratively processes each semantic and vulnerability pair through Specification Generation, \ziqiao{Proof-of-Concept(PoC) Synthesis, PoC Execution,} and Finding Reflection.
Execution failures or specification issues are recorded to guide regeneration, while confirmed vulnerabilities are reported and \ziqiao{potentially} integrated back into the knowledge graph, enabling automated, human-like auditing with continuous refinement and knowledge accumulation.

We evaluate \tool on a dataset of \ziqiao{11 recent Solidity projects from Code4rena, containing 127 contracts and 84 high- and medium-severity vulnerabilities, using a knowledge graph built from 331 historical contests that captures 650 DeFi semantics, 1,226 vulnerability patterns, and 1,954 links between them.
On this dataset, \tool significantly outperforms existing open-source tools, including the industry-leading harness, \textsc{Codex}\xspace by detecting all 21 high-severity vulnerabilities and 90\% of 63 medium-severity vulnerabilities without false positives, demonstrating high precision and recall.
The knowledge graph, assessed by experienced auditors, also reaches high precision with substantial agreement.
In real-world projects, \tool identifies 9 high and 36 medium vulnerabilities, all confirmed and fixed by developers, including severe liquidity-draining issues, while generating concrete proofs of exploitation.
Regarding model sensitivity, \tool demonstrates generalizability across different LLMs, offering a trade-off between detection effectiveness and cost efficiency.}

In summary, this paper makes the following contributions:
\begin{itemize}[leftmargin=15pt]
    \item We propose a novel knowledge-driven, agentic workflow for smart contract vulnerability detection called \tool, which leverages high-level DeFi semantics and vulnerability patterns to capture rich auditing knowledge.
    \item \ziqiao{We construct an auditing knowledge graph from 331 historical contests, comprising 650 DeFi semantics across 13 categories and 1,954 links between these semantics and vulnerability patterns, enabling the generation of more comprehensive auditing specifications.}
    \item Evaluation results show that \tool effectively discovers vulnerabilities, \ziqiao{identifying 21 known high-severity vulnerabilities reported on Code4rena and 9 previously unknown high-severity vulnerabilities in new real-world projects, thereby helping prevent the loss of millions of dollars in user assets.}
\end{itemize}

\section{Motivation}\label{section:motivation}

We present a motivating example to illustrate the challenges of detecting vulnerabilities in diverse smart contracts and our observations to leverage shared DeFi semantics.

\noindent
\textbf{Vulnerabilities often recur in diverse business scenarios.}
Vulnerabilities in smart contracts are recurring, like the notorious reentrancy vulnerabilities causing millions of dollars in loss~\cite{sailfish} and numerous exploits of the forked projects~\cite{zhou2023sok}, due to the inherent design of the EVM~\cite{evm,reentrymaking} and the prevailing code reuse in the Solidity ecosystem~\cite{code-clone-evolve,demystify-clone}.
Besides the recurring vulnerabilities caused by code reuse, distinct smart contracts implementing different business logic may also contain vulnerabilities sharing the same root cause.
To illustrate our observation, we manually study one vulnerability that recurs after 2 years with the same pattern on Code4rena~\cite{c4}, a leading platform for public competitive auditing, from \textit{InsureDAO}\footnote{\url{https://code4rena.com/reports/2022-01-insure}.} in 2022 and \textit{Salty.IO}\footnote{\url{https://code4rena.com/reports/2024-01-saltyio}} in 2024, respectively.
In \textit{InsureDAO}, the first depositor can seed a pool with a tiny amount and then inflate the attribution-to-liquidity ratio, causing later depositors to receive zero or severely undervalued shares and allowing the attacker to drain subsequent deposits.
In \textit{Salty.IO}, the first liquidity provider can likewise manipulate the initial rewards-per-share ratio by adding minimal liquidity and then injecting rewards, which breaks staking-reward accounting and lets the attacker capture rewards that should belong to later participants.
These illustrate that it is difficult to directly map \ziqiao{code or} business scenarios to specific vulnerabilities, creating a gap that hinders automatic tools from generating effective, logic-specific auditing workflows and detection oracles.

\noindent
\textbf{Recurring vulnerabilities often share the same DeFi semantics.}
Figure~\ref{fig:motivation-accounting-pair} presents the two vulnerable functions in \textit{InsureDAO} and \textit{Salty.IO}, respectively. 
Despite differences in business scenarios and implementation, both projects implement a \textit{proportional-shares accounting mechanism for tokens} at a microscopic level, similar to the canonical ERC4626 design~\cite{erc4626}. Let the total shares held by the contract be $\mathbf{S}$, the total assets be $\mathbf{A}$, a user deposit be $a$, and the shares allocated to the user be $s$. The allocation follows the ratio $s = \frac{aS}{A}$.
We refer to such fine-grained functionality that implements a specific economic mechanism as \textit{DeFi semantics}.
As shown in Figure~\ref{fig:motivation-accounting-pair}, \textit{InsureDAO} calculates the expected minted shares $s$ given a specific asset amount \texttt{\_amount}, while \textit{Salty.IO} calculates the assets $a$ corresponding to a given share input \texttt{increaseShareAmount}. Because both projects share the same underlying proportional-shares accounting mechanism, they are vulnerable to the same \textit{first-depositor} attack, where an adversary inflates the ratio $\frac{A}{S}$ to drain liquidity from other users~\cite{uniswapv2,erc4626infaltion}.
By reviewing more similar cases, we observe that vulnerabilities in smart contracts are often associated with particular DeFi semantics that recur across different business scenarios. This observation aligns with prior work~\cite{web3bugs,defiranger,verite,promfuzz,propertygpt}, which shows that understanding high-level DeFi semantics enables the creation of precise invariants or oracles for detecting deep vulnerabilities. Therefore, by constructing an auditing knowledge base bridging low-level code details and high-level DeFi semantics, we can systematically identify potentially related vulnerabilities.

\begin{figure}[t]
    \centering
    \begin{minipage}[t]{0.48\linewidth}
        \begin{minted}[escapeinside=||,xleftmargin=0em,fontsize=\tiny]{solidity}
function addValue(
    uint |\textbf{\_amount}|
) {
    uint |\textbf{\_pool}| = valueAll();
    _attributions = 
    (|\textbf{\_amount}| * |\textbf{totalAttributions}|) / |\textbf{\_pool}|;|\label{line:insuredao-math}|
    attributions[_beneficiary] 
        += _attributions;
}
        \end{minted}
        \centering
        \small InsureDAO (2022)
    \end{minipage}
    \hfill
    \begin{minipage}[t]{0.48\linewidth}
        \begin{minted}[escapeinside=||,xleftmargin=0em,fontsize=\tiny]{solidity}
function _increaseUserShare(
    uint |\textbf{increaseShareAmount}|
) {
    uint virtualRewardsToAdd =
    Math.ceilDiv(
    |\textbf{totalRewards}| * |\textbf{increaseShareAmount}|,
    |\textbf{existingTotalShares}|);|\label{line:saltyio-math}|
    user.virtualRewards
        += uint128(virtualRewardsToAdd);
}
        \end{minted}
        \centering
        \small Salty.IO (2024)
    \end{minipage}
    \caption{Simplified root cause of the recurring ``first depositor'' pattern in \textit{InsureDAO} and \textit{Salty.IO}. The bold key variables involve the proportional share accounting model.}
    \label{fig:motivation-accounting-pair}
\end{figure}


\newcommand*{\repoindex}{\textbf{Repository Index}\xspace}
\newcommand*{\mapper}{\textbf{Knowledge Mapper}\xspace}
\newcommand*{\spec}{\textbf{Specification Generator}\xspace}
\newcommand*{\memory}{\textbf{Working Memory}\xspace}
\newcommand*{\poc}{\textbf{PoC Synthesizer}\xspace}
\newcommand*{\executor}{\textbf{PoC Executor}\xspace}
\newcommand*{\reflector}{\textbf{Finding Reflector}\xspace}

\section{Methodology}\label{section:methodology}

We propose \tool, a novel approach for automated smart contract auditing.

\subsection{Overview}
\begin{figure}
    \centering
    \includegraphics[width=0.98\linewidth]{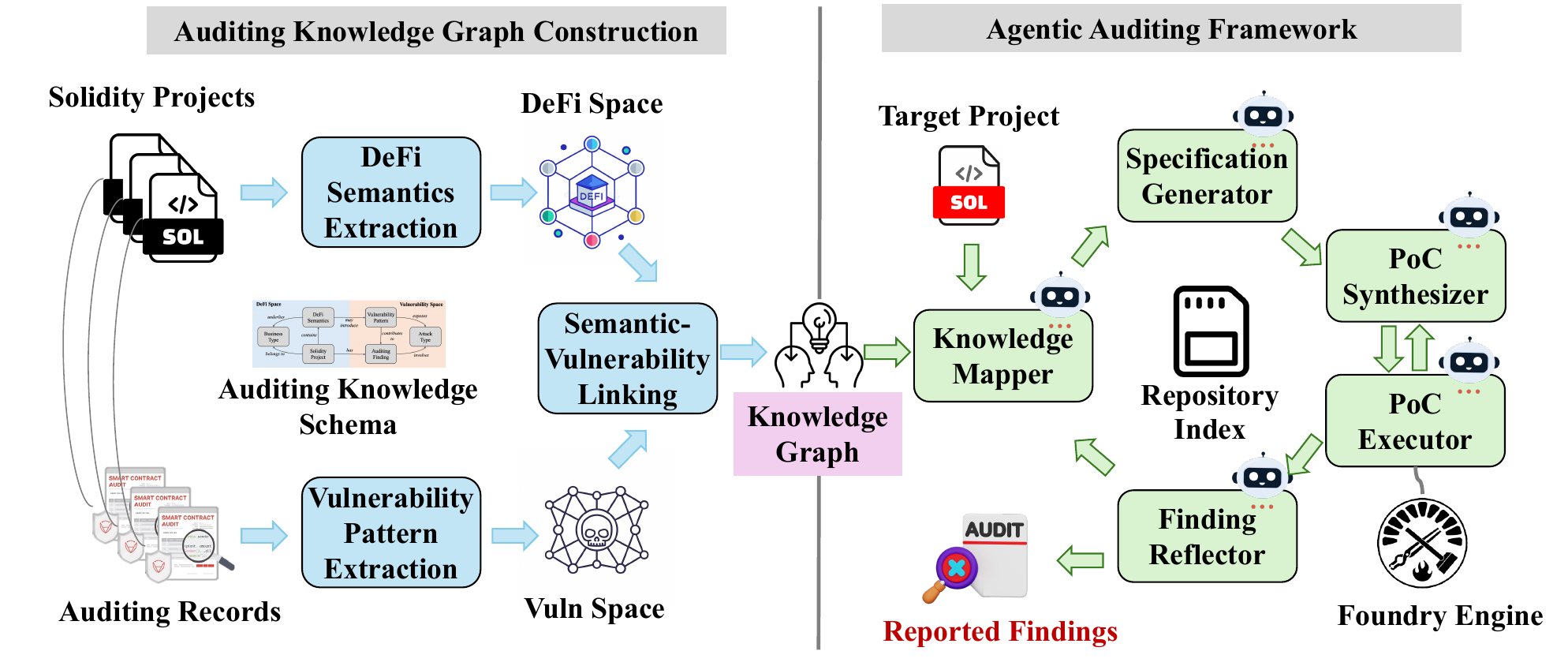}
    \caption{The overview of our method \tool. }
    \label{fig:overview}
\end{figure}

Figure~\ref{fig:overview} provides an overview of \tool, which consists of two main phases: constructing an auditing knowledge graph from historical human reports, and performing knowledge-driven, agentic auditing of new smart contract projects.

We first construct an auditing knowledge graph $\mathcal{G}$ to capture essential smart contract auditing knowledge.
The graph has a bipartite structure: the DeFi Space encodes business types, Solidity projects, and fine-grained DeFi semantics, while the Vulnerability Space captures vulnerability patterns, auditing findings, and attack types.
The two spaces are connected through historical traceability between projects and reports and potential causal links between DeFi semantics and vulnerability patterns. Knowledge is extracted from historical human auditing records using an LLM-based multi-stage pipeline, which abstracts, classifies, and deduplicates DeFi semantics and vulnerability patterns to incrementally build structured, reusable knowledge.

Given a new Solidity project, our auditing framework operates as a multi-agent pipeline centered on a shared \repoindex, which represents the codebase as a structured index and provides agents with efficient code exploration tools \ziqiao{like cross-references for state variables and call-graph traversal}.
The Knowledge Mapper first identifies the project’s business types and retrieves associated DeFi semantics and vulnerability patterns, producing multiple semantic–vulnerability pairs.
\ziqiao{Each pair is then sequentially processed through the Specification Generator, PoC Synthesizer, PoC Executor, and Finding Reflector.}
This loop enables automated, human-like auditing with knowledge accumulation.

\subsection{Auditing Knowledge Graph}\label{sec:distillation}

We first design a graph schema to represent essential auditing knowledge for smart contracts and then extract this knowledge from historical human auditing records.

\subsubsection{Expertise-Inspired Knowledge Schema Design}
Security auditing for smart contracts is a knowledge-intensive process.
To enable comprehensive auditing, we first design a knowledge graph schema inspired by human expertise to model the essential knowledge elements and relationships involved.
Given a smart contract project, an experienced auditor typically analyzes its business scenarios and mentally connects them to relevant economic mechanisms and potential attack surfaces.
For example, if a project's core business involves a \textit{decentralized exchange (DEX)}, it is naturally associated with \textit{token swapping}.
In such cases, an auditor would reasonably suspect potential vulnerabilities such as \textit{price manipulation attacks}, given their prevalence in DEX systems~\cite{defort,gptscan,web3bugs}.
Therefore, associations among key elements are important to provide the necessary knowledge for effective auditing.

\begin{figure}
    \centering
    \includegraphics[width=0.80\linewidth]{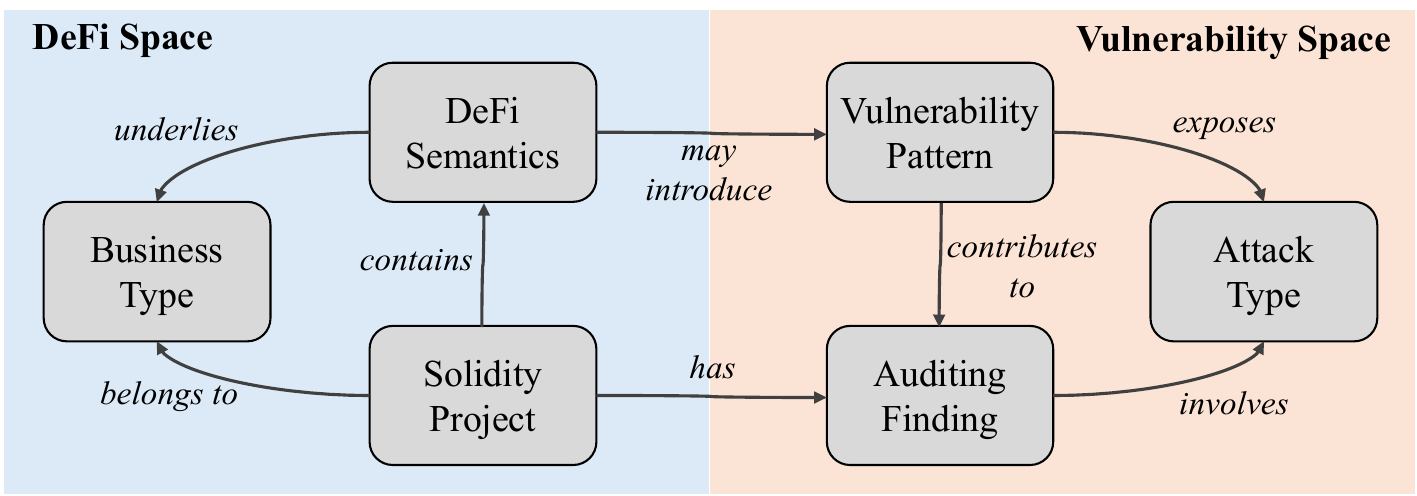}
    \caption{The schema of our auditing knowledge graph.}
    \label{fig:kg}
\end{figure}

The schema of our knowledge graph is illustrated in Figure~\ref{fig:kg}.
It is a bipartite graph consisting of a \textit{DeFi space}, which captures DeFi business scenarios and their associated DeFi semantics, and a \textit{Vulnerability space}, which captures common vulnerability patterns and related attack risks.
These two spaces are connected through historical traceability between Solidity projects and auditing findings, as well as potential causal links between DeFi semantics and vulnerability patterns.
More specifically, the schema consists of the following types of knowledge elements and relationships:

\begin{itemize}[leftmargin=12pt, topsep=3pt, itemsep=2pt, parsep=0pt, partopsep=0pt]
    \item The \textbf{DeFi Space} includes three types of knowledge elements: \textbf{Solidity projects}, which implement specific business logic in DeFi protocols; \textbf{business types} (e.g., \textit{decentralized exchange}), which characterize the high-level functionality of DeFi business logic; and \textbf{DeFi semantics}, which represent fine-grained economic mechanisms (e.g., \textit{token swapping}) implemented in smart contracts. The relationships among these elements include \textit{belongs to} between Solidity projects and business types, \textit{contains} between Solidity projects and DeFi semantics, and \textit{underlies} between DeFi semantics and business types. All three relationships are many-to-many.

    \item The \textbf{Vulnerability Space} also includes three types of knowledge elements: \textbf{auditing findings}, which are reported by human experts and reveal real-world vulnerabilities; \textbf{attack types} (e.g., \textit{Denial-of-Service}), which categorize the associated attack risks; and \textbf{vulnerability patterns}, which summarize recurring vulnerabilities in smart contracts (e.g., \textit{reentrancy could cause inconsistent states}). The relationships among these elements include \textit{contributes to} between vulnerability patterns and auditing findings, \textit{exposes} between vulnerability patterns and attack types, and \textit{involves} between auditing findings and attack types. All three relationships are many-to-many.
    
    \item The connections between the \textbf{DeFi Space} and the \textbf{Vulnerability Space} include a one-to-many \textit{\textbf{has}} relationship between Solidity projects and auditing findings, reflecting their historical traceability, and a many-to-many \textit{\textbf{may introduce}} relationship between DeFi semantics and vulnerability patterns, capturing the potential causal links between them.
\end{itemize}

The business and attack types are predefined based on prior works~\cite{web3bugs,sastsmart}.
The business types include \textit{Lending}, \textit{Dexes}, \textit{Yield}, \textit{Services}, \textit{Derivatives}, \textit{Yield Aggregator}, \textit{Real World Assets}, \textit{Stablecoins}, \textit{Indexes}, \textit{Insurance}, \textit{NFT Marketplace}, \textit{NFT Lending}, and \textit{Cross Chain}, while the attack types include \textit{Access Control}, \textit{Arithmetic}, \textit{Block Manipulation}, \textit{Cryptographic}, \textit{Denial of Service}, \textit{Reentrancy}, and \textit{Storage \& Memory}.

\begin{figure}
    \centering
    \includegraphics[width=0.80\linewidth]{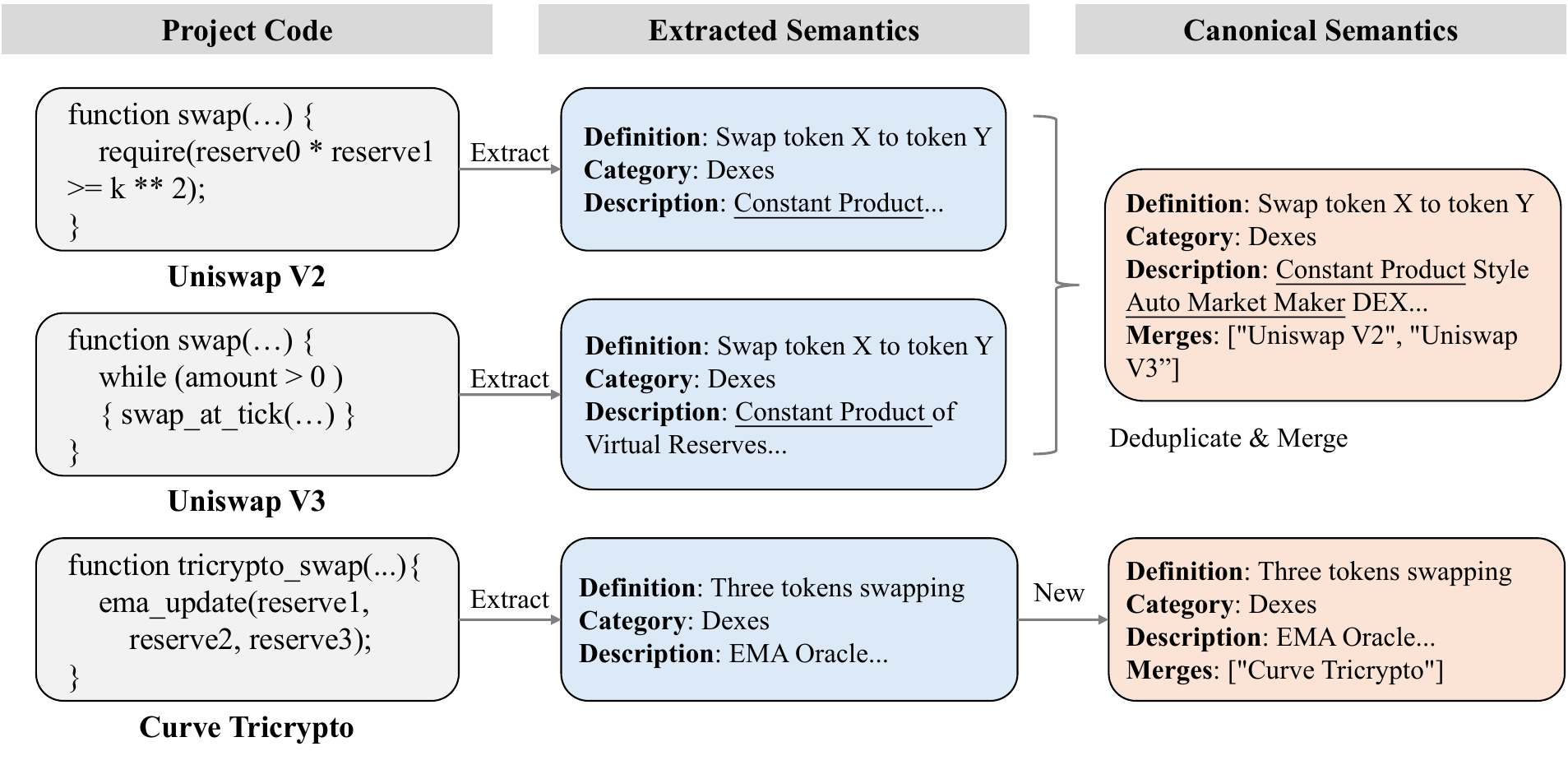}
    \caption{An example for extracting and merging semantics from three DEX projects.}
    \label{fig:semantc-example}
\end{figure}

\subsubsection{LLM-based Knowledge Extraction}\label{sec:knowledge-extract}
Based on the designed schema, we propose an LLM-based multi-stage pipeline to extract knowledge elements and relationships from historical auditing reports written by human experts.
In general, we expect the LLM to mimic human learning by summarizing knowledge into highly abstract, structured representations that serve as reusable experience~\cite{agentlearning}.
We further adopt chain-of-thought~\cite{cot} prompting and in-context learning with examples~\cite{fewcot} to enhance the reasoning reliability of the LLM.

\noindent
\textbf{Stage I: DeFi Semantics Extraction and Abstraction.}
This stage constructs the \textit{DeFi Space} subgraph by traversing and processing a given set of Solidity projects.
We denote the resulting subgraph as $\mathcal{G}^{defi}$.
The detailed steps for processing each project are as follows \ziqiao{and we include the prompts in our artifact~\cite{figshare}}.

\begin{itemize}[leftmargin=12pt, topsep=3pt, itemsep=2pt, parsep=0pt, partopsep=0pt]
\item \textbf{Classifying the project's business type.}
The project is classified into one or more business types.
The LLM evaluates each function or component against predefined type definitions and representative examples.
For instance, a project implementing token swaps following the Uniswap V2 design~\cite{uniswapv2} would be classified as a DEX (decentralized exchange).
For each classification, the LLM provides step-by-step reasoning before producing the final verdict. Following previous work~\cite{gptscan}, a small set of examples for each business type is prepared in the prompt template once to enable in-context learning.
For completeness, we chunk and feed the source code and any documents into the LLM's context window.


\item \textbf{Extracting candidate DeFi semantics.}
The LLM reviews the project's source code and documentation, which are split into chunks to fit within the context window and summarizes each contract's DeFi semantics by abstracting its core mechanics and intentions while removing implementation-specific details.
In addition, we ask the LLM to check if any extracted DeFi semantics can be merged with previous DeFi semantics.
This process captures the fine-grained economic mechanisms embedded in the project.
The project is then added to $\mathcal{G}^{defi}$, and a \textit{belongs to} relationship is established between it and the corresponding business type. If a project belongs to multiple business types, multiple \textit{belongs to} relationships are created accordingly. 


\item \textbf{Deduplicating and summarizing DeFi semantics.}
Once the business type(s) are determined, we retrieve the subset of existing DeFi semantics in $\mathcal{G}^{defi}$ corresponding to the same business type(s).
The LLM compares the newly extracted semantics against this subset. If a candidate overlaps with an existing DeFi semantic, the two are merged and synthesized into a new graph node that captures both variations.
The original node in $\mathcal{G}^{defi}$ is replaced with the merged node, and the current project is linked to it via a \textit{contains} relationship.
Otherwise, the candidate is added to $\mathcal{G}^{defi}$ as a new node, and \textit{contains} and \textit{underlies} relationships are established with the project and its business type(s), respectively, following the schema.
This combined deduplication and abstraction step reduces redundancy and efficiently builds a structured, reusable knowledge graph.
\end{itemize}

We present an illustrative example of the workflow in Figure~\ref{fig:semantc-example} that extracts and merges DeFi semantics from three distinct DEX projects, UniswapV2~\cite{uniswapv2}, UniswapV3~\cite{uniswapv3}, and Curve Tricrypto~\cite{curve}.
For UniswapV2 and UniswapV3, although their implementations are dramatically different, they share the same DeFi semantics \textit{Swap token X to token Y}, and we summarize their economic model \textit{Constant Product} into the merged semantic node.
However, Curve Tricrypto supports swapping with three tokens, unlike the two-token swapping model of Uniswap families.
Therefore, we consider it a distinct DeFi semantic node.

This stage begins with an empty $\mathcal{G}^{defi}$. As more projects are processed, the graph is continuously updated, and knowledge about DeFi semantics accumulates over time.

\noindent
\textbf{Stage II: Vulnerability Pattern Extraction and Summarization.}
This stage constructs the \textit{Vulnerability Space} subgraph by processing a set of human auditing reports associated with the analyzed projects.
We denote the resulting subgraph as $\mathcal{G}^{vuln}$.
The overall workflow of constructing the subgraph follows the same \textit{extracting–classifying–deduplicating} paradigm and similar prompt templates as Stage I, \ziqiao{as presented in our artifact~\cite{figshare}}, but operates on auditing reports instead of source code and focuses on vulnerability-related knowledge.

\begin{itemize}[leftmargin=12pt, topsep=3pt, itemsep=2pt, parsep=0pt, partopsep=0pt]

\item \textbf{Classifying the report's attack type.}
Each finding is classified into one or more attack types (e.g., \textit{Reentrancy}) using the classification prompt with predefined categories and examples.

\item \textbf{Extracting candidate vulnerability patterns.}
Specifically, the LLM abstracts each auditing finding into a vulnerability pattern by capturing its root cause and core logic, rather than implementation details. 

\item \textbf{Deduplicating and summarizing vulnerability patterns.}
During deduplication, the extracted patterns are compared against existing nodes in $\mathcal{G}^{vuln}$ within the same attack type(s). Overlapping patterns are merged and refined into unified representations, while novel patterns are added as new nodes. Correspondingly, \textit{contributes to} relationships are established between auditing findings and vulnerability patterns, and \textit{involves} relationships are created between findings and attack types, following the schema definition.
\end{itemize}

This stage also begins with an empty $\mathcal{G}^{vuln}$ and incrementally accumulates structured knowledge about vulnerability patterns as more auditing reports are processed.

\noindent
\textbf{Stage III: Causal Linking between DeFi Semantics and Vulnerability Patterns.}
Given the two subgraphs $\mathcal{G}^{defi}$ and $\mathcal{G}^{vuln}$, we further establish causal links between DeFi semantics and vulnerability patterns.

\ziqiao{Specifically, we split the two subgraphs into a set of DeFi-semantic chunks $\mathcal{D}$ and a set of vulnerability-pattern chunks $\mathcal{V}$.
We then prompt the LLM to emit all links over the Cartesian product $\mathcal{D}\times\mathcal{V}$, i.e., for every chunk pair $(d,v)$ with $d\in\mathcal{D}$ and $v\in\mathcal{V}$.
A link is established between a DeFi semantic and a vulnerability pattern if the semantic can potentially introduce or contribute to the vulnerability characterized by the pattern.}\ziqiao{We include the prompt in our artifact~\cite{figshare}.}

\ziqiao{After processing all chunks from $\mathcal{D}$ and $\mathcal{V}$, we obtain the final auditing knowledge graph $\mathcal{G}$.}

\subsection{Agentic Auditing Framework}

Given a new Solidity project, we employ an agentic framework consisting of four LLM-based components and \ziqiao{a PoC executor}, which leverages the constructed knowledge graph to perform an iterative exploiting and verification loop.

\subsubsection{Overall Workflow}

Given a new Solidity project, the multi-agent pipeline proceeds as follows. \tool first constructs a global \repoindex that captures the call graph, state-variable reference graph, contract inheritance graph, and a source code index organized by contracts and functions. The \repoindex is exposed to agents through dedicated tools, enabling efficient code exploration while reducing token consumption.
The \mapper identifies the project's business types and retrieves the associated DeFi semantics and linked vulnerability patterns from $\mathcal{G}$, producing multiple semantic–vulnerability pairs.
Each pair is processed sequentially through the \spec, which concretizes abstract knowledge into an auditing specification including several key states of the attacking scenarios; the \poc, which \ziqiao{writes a proof-of-concept} based on the specification; the \executor, which runs the proof-of-concept and collects coverage and execution trace; and the \reflector, which validates violations.
\ziqiao{Execution failures in the \executor are recorded and trigger regeneration of the PoC.}
\ziqiao{If a violation from \spec stems from a problematic specification, the workflow returns to the \spec to regenerate it; if a valid vulnerability is confirmed, it is reported with a concrete proof-of-concept, title, description, and related functions.
This loop iterates over all pairs, enabling iterative refinement and human-like auditing.}

\subsubsection{Reliability Assurance Mechanisms}
\chong{Many agents in \tool are required to produce complex (semi-)structured outputs, which are difficult to generate accurately in a single attempt. To address this challenge, we introduce an iterative \textbf{Drafting} mechanism. For each agent, \tool disallows direct generation of the final output and instead provides a set of dedicated \textit{updateX} tools, together with a tailored system prompt that encourages iterative reflection and refinement of individual fields before committing the final output. For example, the \spec is equipped with tools such as \textit{updatePreStateContract} and \textit{updatePreStateVariable}, allowing it to progressively refine different fields of an auditing specification.
Furthermore, inspired by \cite{sweagent}, \tool integrates \textbf{Guardrails} via \repoindex into \textit{updateX} tools that update code-related fields, reducing hallucinations by rejecting forged names and enforcing consistency with the underlying codebase.}


\subsubsection{Knowledge Mapper}\label{sec:workflow-knowledge-mapper}
We first analyze the given project to identify its business types and extract the DeFi semantics involved by prompting the LLM, \ziqiao{reusing the same process as defined in Section~\ref{sec:knowledge-extract}.}
For each extracted DeFi semantic, we prompt the LLM to identify its matches among the semantics associated with the identified business types in the knowledge graph $\mathcal{G}$.
We then retrieve the linked vulnerability patterns and assemble a set of semantic–vulnerability pairs.
Next, we will iterate over every semantic-vulnerability pair for the following procedures.

\subsubsection{Specification Generator}
We then concretize the mapped \textit{abstract} knowledge of DeFi semantics and associated vulnerability patterns into project-specific, \textit{actionable} auditing specifications \ziqiao{by locating the corresponding functions or contracts.}

\noindent\textbf{Definition of Auditing Specification.} \ziqiao{We first define \textit{State} under the context of audit specifications: \textit{The expected contracts to be deployed and the invariants of the state variables at a given time.} The schema of an audit specification is three key states: \textit{Initial State}, \textit{Pre-Vuln State}, \textit{Post-Vuln State}, and a transaction sequence.}
\ziqiao{Each \textit{State} shares the same definition, and the transaction sequence encodes the key contract functions' interactions that modify the related state variables.}
Each semantic-vulnerability pair is supposed to have a unique auditing specification \ziqiao{on a given project}.
\ziqiao{We take testing a \textit{Dexes} project for a typical \textit{Price Manipulation} vulnerability as an example to illustrate the contents of audit specifications.}

\begin{itemize}[leftmargin=12pt, topsep=3pt, itemsep=2pt, parsep=0pt, partopsep=0pt]

\item \textbf{Initial State.} The state usually contains the expected pools and tokens that should be available and funded for testing.

\item \textbf{Pre-Vuln State.} The pool is supposed to hold some liquidity and not paused.

\item \textbf{Post-Vuln State.} The liquidity is drained, and the price is unreasonably shifted.

\item \ziqiao{\textbf{Transaction Sequence.} Usually, a sequence of operations like transferring and swapping tokens.}

\end{itemize}


\subsubsection{PoC Synthesizer} We synthesize PoCs in the form of Foundry~\cite{foundry} test cases from the generated specifications.
Specifically, we prompt the LLM to encode \textit{Initial State} to a \textit{Foundry setUp} function that deploys contracts and initializes accounts, translate \textit{Pre-Vuln State} and \textit{Post-Vuln State} invariants into \textit{require} statements as oracles.
The agent will synthesize the test until it builds.

\subsubsection{PoC Executor} We run the PoCs for the configured timeout and collect the line coverage, state changes and full trace.
\ziqiao{The line coverage and trace are intended to assess the PoC quality. If a PoC does not exercise a minimum proportion of lines or does not touch the state variables defined in the audit specification, we drop the PoC immediately. This serves as a cheap validation that drops low-quality PoC without wasting tokens.}

\subsubsection{Finding Reflector}\label{sec:finding-reflector} Once \tool finds a violation, we first prompt LLM to compare the state changes and trace with the audit specification to see if the violation matches the vulnerability pattern.
\ziqiao{Specifically, we drive the reflector to read through call chains of related functions, including their comments, to understand the implicit assumptions of the projects.}
If the violation fails to match the specification, \tool further classifies it into:

\begin{itemize}[leftmargin=12pt, topsep=3pt, itemsep=2pt, parsep=0pt, partopsep=0pt]
\item \textbf{Expected Behavior.} Many violations do not indicate vulnerabilities because smart contracts are often designed to revert under certain conditions. For example, the common \textit{onlyOwner} check ensures that only privileged users can execute a function; any abort caused by this check is expected and not a valid vulnerability.

\item \textbf{Problematic Specification or PoC.} Some violations arise from incomplete or inaccurate PoCs or specifications, such as missing deployment setup or incorrect assumptions about initial states. For instance, failing to initialize contracts may produce false alarms for oracles.
\end{itemize}

\ziqiao{For both cases, \tool triggers another attempt for \spec and \poc with the code and reasons of the failure.}

Even if the vulnerability matches the specification, \tool performs a further review if the vulnerability is out-of-scope according to the project README and Code4rena general rules\footnote{\url{https://docs.code4rena.com/bounties/bounty-criteria}}.
For example, many projects state they will not integrate any fee-on-transfer tokens, and thus \tool will rule out findings involving such tokens accordingly, even though the findings could be valid.
Once the finding passes the review, \tool will report it as a true finding.
\ziqiao{\tool also supports overriding the review rules.}

\newcommand*{\train}{$\mathbf{DS}_{T}$\xspace}
\newcommand*{\eval}{$\mathbf{AuditEval}$\xspace}

\newcommand*{\promfuzz}{\textsc{PromFuzz}\xspace}
\newcommand*{\propertygpt}{\textsc{PropertyGPT}\xspace}
\newcommand*{\llmaudit}{\textsc{LLM-SmartAudit}\xspace}
\newcommand*{\smartinv}{\textsc{SmartInv}\xspace}
\newcommand*{\codex}{\textsc{Codex}\xspace}
\newcommand*{\codexevm}{\textsc{Codex-EVM}\xspace}
\newcommand*{\evmbench}{\textsc{EVMBench}\xspace}
\newcommand*{\gptscan}{\textsc{GPTScan}\xspace}
\newcommand*{\wokg}{\textsc{\textit{w/o} KG}\xspace}
\newcommand*{\wcodex}{\textsc{\textit{w/} Codex}\xspace}
\newcommand*{\toolnkg}{$\tool_{\wokg}$\xspace}
\newcommand*{\toolcodex}{$\tool_{\wcodex}$\xspace}

\newcommand*{\nkg}{\textsc{NKG}\xspace}

\newcommand*{\ityfuzz}{\textsc{ItyFuzz}\xspace}
\newcommand*{\gptfivefour}{\textsc{GPT-5.4}\xspace}
\newcommand*{\gptfive}{\gptfivefour}
\newcommand*{\dpskpro}{\textsc{DeepSeek-V4-Pro}\xspace}
\newcommand*{\gptfivemini}{\textsc{GPT-5.4-mini}\xspace}

\newcommand*{\high}{\textit{High}\xspace}
\newcommand*{\medium}{\textit{Medium}\xspace}
\newcommand*{\low}{\textit{Low}\xspace}
\newcommand*{\qa}{\textit{Quality Assurance}\xspace}




\section{Evaluation}\label{section:evaluation}
We conduct extensive experiments to evaluate the effectiveness of \tool for smart contract auditing. Specifically, we aim to answer the following research questions:

\begin{itemize}[leftmargin=12pt, topsep=3pt, itemsep=2pt, parsep=0pt, partopsep=0pt]
    \item \textbf{RQ1 (Dataset-based Evaluation)}: How does \tool compare with existing tools in end-to-end auditing on a dataset constructed from historical auditing reports?
    \item \textbf{RQ2 (Ablation Study)}: To what extent do the knowledge graph and the agentic framework contribute to the overall performance of \tool? What is the quality of the constructed knowledge graph?
    \item \textbf{RQ3 (Unknown Vulnerability Discovery)}: How effective is \tool at discovering previously unknown vulnerabilities in newly developed real-world projects?
    \item \textbf{RQ4 (Model Sensitivity Analysis)}: How sensitive is \tool to different foundation models in terms of detection effectiveness and cost efficiency?
\end{itemize}

\subsection{Experimental Setup}

\noindent \textbf{Implementation.} \tool is written in around \textit{54.2k} lines of Rust code and relies on Foundry~\cite{foundry} as the PoC execution engine. All agents in \tool are equipped with context window management and auto conversation compaction to handle various real-world auditing scenarios. \ziqiao{\tool by default prompts \gptfivefour to balance the cost, output quality, and inference speed, and uses all default parameters. We also include an experiment for model sensitivity in Section~\ref{sec:eval-model-sensitivity}.}

\noindent \textbf{Data Collection.} We collect public Solidity audit contests from Code4rena~\cite{c4}, which provides free access to source code and manually reviewed, deduplicated, and triaged audit reports. To avoid data leakage, we split the contests according to the knowledge cutoff date of \gptfivefour  (\textit{31/08/2025}): contests released before this date are used to construct the knowledge graph, while those released during the subsequent 6 months \ziqiao{without filtering} are reserved for evaluating \tool.

\noindent
\textbf{Knowledge Graph.} A total of 331 contests are used to construct the knowledge graph with \gptfivefour.
The resulting graph contains 650 deduplicated DeFi semantics consolidated from 1,776 extracted candidates, 1,226 vulnerability patterns derived from 3,251 audit findings, and 1,954 validated links between semantics and patterns. Building the graph incurs a one-time token cost of \$1,336, corresponding to an average cost of \$4.0 per Solidity \wanxu{contest}.

\noindent
\textbf{Evaluation Dataset.} We construct \eval, the evaluation dataset, using all Solidity projects collected during the data collection phase without additional filtering.
For each project, we use the corresponding audit reports as ground truth, which contain manually verified vulnerabilities categorized into three severity levels: \high, \medium, and \qa.
Following prior works~\cite{llmaudit,web3bugs}, we exclude all \qa issues, as they primarily concern code style or gas optimization and do not affect contract security.
As summarized in Table~\ref{tab:project_info}, \eval comprises 11 projects with 21 high-severity and 63 medium-severity vulnerabilities.
In total, the dataset comprises 127 smart contracts and approximately 58K lines of code, ranging from small projects with 1.2K lines of code to large projects with over 12K lines of code.

\begin{table}[t]
\centering
\footnotesize
\caption{Summary of our evaluation dataset \eval.}
\label{tab:project_info}
\resizebox{\columnwidth}{!}{%
\begin{tabular}{l|lrrrr}
\toprule
ID & Project  & \#Contract & \#LoC & \#\high & \#\medium \\
\midrule
\rowcolor{gray!12} Garden   & Garden               & 6  & 1,775  & 0  & 1  \\
Sequence & Sequence             & 21 & 4,630  & 2  & 4  \\
\rowcolor{gray!12} Hybra    & Hybra Finance        & 14 & 6,087  & 1  & 9  \\
Ekubo    & Ekubo                & 29 & 10,971 & 0  & 4  \\
\rowcolor{gray!12} Megapot  & Megapot              & 6  & 4,412  & 3  & 8  \\
SukukFi  & SukukFi              & 4  & 5,057  & 1  & 3  \\
\rowcolor{gray!12} Merkl    & Merkl                & 2  & 1,226  & 0  & 3  \\
Panoptic & Panoptic: Next Core  & 5  & 12,796 & 3  & 19 \\
\rowcolor{gray!12} Olas-Gov & Olas-Governance      & 15 & 1,932  & 0  & 1  \\
Olas-Reg & Olas-Registries      & 8  & 2,834  & 5  & 2  \\
\rowcolor{gray!12} Olas-Tok & Olas-Tokenomics      & 17 & 6,090  & 6  & 9  \\
\hline
\textbf{Sum} & - & 127 & 57,810 & 21 & 63 \\
\bottomrule
\end{tabular}}
\end{table}

\noindent \textbf{Baseline Methods.} We evaluate \tool against several representative open-source LLM-based baselines spanning different categories. We exclude approaches that depend on significant human intervention and thus do not enable fully automated end-to-end auditing, such as methods requiring manually written specifications, vulnerability oracles, or auditing rules. The selected baselines are as follows:
\begin{itemize}[leftmargin=12pt, topsep=3pt, itemsep=2pt, parsep=0pt, partopsep=0pt]
    \item \textbf{LLM-based Invariant Inference.}
    \propertygpt~\cite{propertygpt} and \smartinv~\cite{smartinv} leverage LLMs to infer properties or invariants and check them with a verification engine: \propertygpt pairs retrieval-augmented properties from Certora~\cite{certora} with a source-level symbolic-execution prover, whereas \smartinv generates invariants with a fine-tuned LLaMA model and verifies them via proof-of-correctness backed by a bounded model checker. We exclude \promfuzz~\cite{promfuzz}, another method in this category, because it produces no findings on \eval, \ziqiao{though we have fully evaluated it}.

    \item \textbf{LLM-assisted Static Auditing.}
    \llmaudit~\cite{llmaudit} and \gptscan~\cite{gptscan} read contract code and emit
    vulnerability findings directly, without synthesizing formal specifications. \llmaudit is
    a pure LLM auditor, whereas \gptscan couples the LLM with static analysis, using GPT for
    scenario/property matching and static confirmation (e.g., data-flow tracing and static
    symbolic execution) to suppress false positives.

    \item \textbf{Coding Agent Harness.}
    \codex~\cite{codex} (\ziqiao{version 0.136.0 used in our experiments}) is an industry-leading LLM coding agent harness. Although mainly designed
    for coding, a recent study \evmbench~\cite{evmbench} shows that \codex is capable of
    complex auditing with reproducible proofs-of-concept. We denote \codexevm for \codex using
    the \evmbench prompts.
\end{itemize}

Additionally, to ablate the contributions of the knowledge graph and our agentic auditing framework, we derive two variants of \tool:
\begin{itemize}[leftmargin=12pt, topsep=3pt, itemsep=2pt, parsep=0pt, partopsep=0pt]
    \item \toolnkg removes the use of the knowledge graph (i.e., the Knowledge Mapper) from \tool's agentic workflow and instead directly prompts the LLM to autonomously explore the codebase and generate auditing specifications. This variant isolates the contribution of the knowledge graph.

    \item \toolcodex replaces the entire agentic workflow of \tool with \codexevm while preserving access to \tool's knowledge graph through a knowledge retrieval skill implemented using the \textit{bge-m3}~\cite{bgem3} embedding model. This variant isolates the contribution of \tool's agentic framework by comparing it against an advanced agent harness under the same knowledge graph.
\end{itemize}

\noindent
\textbf{End-to-End Evaluation Environment.} We run all baselines using \gptfivefour, the same LLM as \tool.
To ensure a consistent evaluation environment, we build a shared Docker image containing the \eval projects, the evaluated tools, and all required dependencies, including the Solidity compiler, Foundry, and Hardhat, so that every experiment is executed with identical inputs and a reproducible software environment.
Each tool is executed in an isolated container provisioned with 16 CPU cores, two RTX A5000 GPUs (used for embedding computation in \toolcodex and fine-tuned model inference in \smartinv), and 32GiB of memory. For \llmaudit, we enable both the Targeted Analysis (TA) and Broad Analysis (BA) modes, and count the findings from both modes together. For \codexevm and \toolcodex, we enable all available permissions and tools while restricting container networking to inference endpoints only, preventing access to external resources and eliminating potential data leakage.
Like other fuzzing-based approaches, \tool continuously explores potential vulnerabilities until the allocated token budget is exhausted. For a fair comparison, each tool is allowed to consume up to \$200 worth of LLM tokens per project. We repeat every experiment five times to mitigate the non-determinism of LLM outputs and aggregate all unique findings across runs.

\subsection{RQ1: Dataset-based Evaluation}\label{sec:rq1}
We run \tool and all baselines on \eval and compare their reported findings against the ground truth. For each tool, we report the numbers of correctly identified vulnerabilities (\high and \medium) and false positives. Our artifact~\cite{figshare} contains several case studies on \eval to illustrate the effectiveness of \tool.

\noindent
\textbf{Results.} Table~\ref{tab:rq1-bug-finding} presents the evaluation results. \tool significantly outperforms all baselines in automated smart contract auditing, identifying all 21 \high vulnerabilities (100\%) and 57 of the 63 \medium vulnerabilities (90.5\%) without producing any false positives. In contrast, the baselines detect substantially fewer vulnerabilities, ranging from 1 to 24, while generating considerably more false positives. Furthermore, on eight projects, \tool successfully identifies all ground-truth vulnerabilities with zero false positives. The high precision and recall demonstrate the strong practical potential of \tool for end-to-end smart contract auditing.

\newcommand*{\pfuzz}{\textsc{P-Fuzz}\xspace}
\newcommand*{\pgpt}{\textsc{P-GPT}\xspace}
\newcommand*{\laudit}{\textsc{L-Audit}\xspace}
\newcommand*{\sinv}{\textsc{S-Inv}\xspace}
\newcommand*{\ckg}{\textsc{C-KG}\xspace}
\newcommand*{\ceb}{\textsc{C-EB}\xspace}
\newcommand*{\gscan}{\textsc{G-Scan}\xspace}

\begin{table*}[t]
    \centering
    \footnotesize
    \setlength{\tabcolsep}{4pt}
    \caption{Results of the evaluation on \eval. We highlight in \textbf{bold} the results that achieve complete coverage of all ground-truth vulnerabilities.}
    \label{tab:rq1-bug-finding}
    \renewcommand{\arraystretch}{1.05}
    \begin{tabular}{l|c|cccccccc}
        \toprule
        \multicolumn{1}{c|}{\multirow{2}{*}{\textbf{Project}}} & \multirow{2}{*}{\begin{tabular}[c]{@{}c@{}}\#Ground-Truth\\ (\high\thinspace/\thinspace \medium)\end{tabular}} & \multicolumn{8}{c}{\textbf{\#Reported Findings} (True \high\thinspace/\thinspace True \medium\thinspace/\thinspace False Positive)}                                                                                                                                                                                                                                                                                                                                                  \\ \cline{3-10} 
        \multicolumn{1}{c|}{\rule{0pt}{2.4ex}}                                  &                                                                                                                & \propertygpt                                     & \smartinv                                        & \gptscan                                         & \llmaudit                                                & \multicolumn{1}{c|}{\codexevm}                                                 & \tool                                                      & \wokg                                                 & \wcodex                                               \\ 
        \midrule
        \rowcolor{gray!12} Garden                              & 0\thinspace/\thinspace 1                                                                                       & 0\thinspace/\thinspace 0\thinspace/\thinspace 0  & 0\thinspace/\thinspace 0\thinspace/\thinspace 0  & 0\thinspace/\thinspace 0\thinspace/\thinspace 0  & 0\thinspace/\thinspace 0\thinspace/\thinspace 4          & \multicolumn{1}{c|}{0\thinspace/\thinspace 0\thinspace/\thinspace 1}          & 0\thinspace/\thinspace \textbf{1}\thinspace/\thinspace 0   & 0\thinspace/\thinspace \textbf{1}\thinspace/\thinspace 0 & 0\thinspace/\thinspace 0\thinspace/\thinspace 1          \\
        Sequence                                               & 2\thinspace/\thinspace 4                                                                                       & 0\thinspace/\thinspace 0\thinspace/\thinspace 0  & 0\thinspace/\thinspace 0\thinspace/\thinspace 4  & 0\thinspace/\thinspace 0\thinspace/\thinspace 0  & 0\thinspace/\thinspace 0\thinspace/\thinspace 5          & \multicolumn{1}{c|}{0\thinspace/\thinspace 1\thinspace/\thinspace 1}          & \textbf{2\thinspace/\thinspace 4}\thinspace/\thinspace 0   & 0\thinspace/\thinspace 0\thinspace/\thinspace 0          & 1\thinspace/\thinspace 1\thinspace/\thinspace 1          \\
        \rowcolor{gray!12} Hybra                               & 1\thinspace/\thinspace 9                                                                                       & 0\thinspace/\thinspace 2\thinspace/\thinspace 13 & 0\thinspace/\thinspace 0\thinspace/\thinspace 6  & 0\thinspace/\thinspace 1\thinspace/\thinspace 9  & \textbf{1}\thinspace/\thinspace 3\thinspace/\thinspace 3 & \multicolumn{1}{c|}{\textbf{1}\thinspace/\thinspace 0\thinspace/\thinspace 0} & \textbf{1}\thinspace/\thinspace 7\thinspace/\thinspace 0   & \textbf{1}\thinspace/\thinspace 2\thinspace/\thinspace 0 & \textbf{1}\thinspace/\thinspace 0\thinspace/\thinspace 0 \\
        Ekubo                                                  & 0\thinspace/\thinspace 4                                                                                       & 0\thinspace/\thinspace 0\thinspace/\thinspace 0  & 0\thinspace/\thinspace 0\thinspace/\thinspace 1  & 0\thinspace/\thinspace 0\thinspace/\thinspace 2  & 0\thinspace/\thinspace 0\thinspace/\thinspace 4          & \multicolumn{1}{c|}{0\thinspace/\thinspace 0\thinspace/\thinspace 2}          & 0\thinspace/\thinspace \textbf{4}\thinspace/\thinspace 0   & 0\thinspace/\thinspace 0\thinspace/\thinspace 0          & 0\thinspace/\thinspace 0\thinspace/\thinspace 2          \\
        \rowcolor{gray!12} Megapot                             & 3\thinspace/\thinspace 8                                                                                       & 0\thinspace/\thinspace 1\thinspace/\thinspace 0  & 0\thinspace/\thinspace 0\thinspace/\thinspace 3  & 1\thinspace/\thinspace 0\thinspace/\thinspace 0  & 2\thinspace/\thinspace 2\thinspace/\thinspace 4          & \multicolumn{1}{c|}{0\thinspace/\thinspace 3\thinspace/\thinspace 0}          & \textbf{3}\thinspace/\thinspace 6\thinspace/\thinspace 0   & 1\thinspace/\thinspace 3\thinspace/\thinspace 0          & 0\thinspace/\thinspace 2\thinspace/\thinspace 1          \\
        SukukFi                                                & 1\thinspace/\thinspace 3                                                                                       & 0\thinspace/\thinspace 0\thinspace/\thinspace 0  & 0\thinspace/\thinspace 0\thinspace/\thinspace 3  & 0\thinspace/\thinspace 0\thinspace/\thinspace 0  & 0\thinspace/\thinspace 1\thinspace/\thinspace 3          & \multicolumn{1}{c|}{\textbf{1}\thinspace/\thinspace 1\thinspace/\thinspace 0} & \textbf{1\thinspace/\thinspace 3}\thinspace/\thinspace 0   & \textbf{1}\thinspace/\thinspace 2\thinspace/\thinspace 0 & \textbf{1}\thinspace/\thinspace 1\thinspace/\thinspace 0 \\
        \rowcolor{gray!12} Merkl                               & 0\thinspace/\thinspace 3                                                                                       & 0\thinspace/\thinspace 1\thinspace/\thinspace 2  & 0\thinspace/\thinspace 0\thinspace/\thinspace 2  & 0\thinspace/\thinspace 0\thinspace/\thinspace 2  & 0\thinspace/\thinspace 2\thinspace/\thinspace 0          & \multicolumn{1}{c|}{0\thinspace/\thinspace 0\thinspace/\thinspace 0}          & 0\thinspace/\thinspace \textbf{3}\thinspace/\thinspace 0   & 0\thinspace/\thinspace 1\thinspace/\thinspace 0          & 0\thinspace/\thinspace 1\thinspace/\thinspace 0          \\
        Panoptic                                               & 3\thinspace/\thinspace 19                                                                                      & 1\thinspace/\thinspace 0\thinspace/\thinspace 0  & 0\thinspace/\thinspace 0\thinspace/\thinspace 4  & 0\thinspace/\thinspace 0\thinspace/\thinspace 9  & 1\thinspace/\thinspace 2\thinspace/\thinspace 5          & \multicolumn{1}{c|}{1\thinspace/\thinspace 1\thinspace/\thinspace 2}          & \textbf{3\thinspace/\thinspace 19}\thinspace/\thinspace 0  & 1\thinspace/\thinspace 0\thinspace/\thinspace 0          & 2\thinspace/\thinspace 0\thinspace/\thinspace 0          \\
        \rowcolor{gray!12} Olas-Gov                            & 0\thinspace/\thinspace 1                                                                                       & 0\thinspace/\thinspace 0\thinspace/\thinspace 3  & 0\thinspace/\thinspace 0\thinspace/\thinspace 0  & 0\thinspace/\thinspace 0\thinspace/\thinspace 1  & 0\thinspace/\thinspace 0\thinspace/\thinspace 4          & \multicolumn{1}{c|}{0\thinspace/\thinspace \textbf{1}\thinspace/\thinspace 0} & 0\thinspace/\thinspace \textbf{1}\thinspace/\thinspace 0   & 0\thinspace/\thinspace 0\thinspace/\thinspace 0          & 0\thinspace/\thinspace \textbf{1}\thinspace/\thinspace 0 \\
        Olas-Reg                                               & 5\thinspace/\thinspace 2                                                                                       & 0\thinspace/\thinspace 0\thinspace/\thinspace 1  & 0\thinspace/\thinspace 0\thinspace/\thinspace 2  & 0\thinspace/\thinspace 0\thinspace/\thinspace 0  & 2\thinspace/\thinspace 0\thinspace/\thinspace 5          & \multicolumn{1}{c|}{0\thinspace/\thinspace 0\thinspace/\thinspace 0}          & \textbf{5\thinspace/\thinspace 2}\thinspace/\thinspace 0   & 0\thinspace/\thinspace 0\thinspace/\thinspace 0          & 3\thinspace/\thinspace 0\thinspace/\thinspace 0          \\
        \rowcolor{gray!12} Olas-Tok                            & 6\thinspace/\thinspace 9                                                                                       & 0\thinspace/\thinspace 0\thinspace/\thinspace 0  & 0\thinspace/\thinspace 1\thinspace/\thinspace 4  & 3\thinspace/\thinspace 1\thinspace/\thinspace 7  & \textbf{6}\thinspace/\thinspace 2\thinspace/\thinspace 6 & \multicolumn{1}{c|}{1\thinspace/\thinspace 0\thinspace/\thinspace 0}          & \textbf{6}\thinspace/\thinspace 7\thinspace/\thinspace 0   & 1\thinspace/\thinspace 1\thinspace/\thinspace 0          & 2\thinspace/\thinspace 0\thinspace/\thinspace 0          \\ 
        \midrule
        \textbf{Sum}                                           & 21\thinspace/\thinspace 63                                                                                     & 1\thinspace/\thinspace 4\thinspace/\thinspace 19 & 0\thinspace/\thinspace 1\thinspace/\thinspace 29 & 4\thinspace/\thinspace 2\thinspace/\thinspace 30 & 12\thinspace/\thinspace 12\thinspace/\thinspace 43       & \multicolumn{1}{c|}{4\thinspace/\thinspace 7\thinspace/\thinspace 6}          & \textbf{21}\thinspace/\thinspace 57\thinspace/\thinspace 0 & 5\thinspace/\thinspace 10\thinspace/\thinspace 0         & 10\thinspace/\thinspace 6\thinspace/\thinspace 5         \\ 
        \bottomrule
    \end{tabular}
\end{table*}

\noindent
\textbf{Compared to LLM-based Invariant Inference.} 
\chong{\propertygpt and \smartinv perform poorly on \eval, detecting only 5 (1 \high + 4 \medium) and 1 \medium vulnerabilities, respectively, while producing 19 and 29 false positives.
Their limitations stem from three fundamental differences from \tool.
\textit{First}, their auditing knowledge is closed and static, bound to a corpus of \emph{manually written} Certora properties (\propertygpt), or a labeled fine-tuning distribution (\smartinv), \ziqiao{or limited templates (\promfuzz),} and thus generalizes poorly to broader projects.
In contrast, \tool mines and leverages an extensible knowledge base that can map abstract DeFi semantics and vulnerability patterns to concrete functions and contracts.
\textit{Second}, they rely on standalone invariants at the function or contract level, which cannot capture the cross-contract, multi-step interactions underlying real-world DeFi exploits.
Instead, \tool models an auditing specification as an \emph{exploit-oriented} attack scenario consisting of \textit{Initial}, \textit{Pre-Vuln}, and \textit{Post-Vuln} states together with a transaction sequence encoding inter-contract interactions and deployment conditions. \textit{Third}, \tool validates each finding with an executable Foundry PoC: the \textit{Initial State} is instantiated in \textit{setUp}, and the \textit{Pre-/Post-Vuln} states become \textit{require} oracles, ensuring that every reported vulnerability is witnessed by a concrete trace.}

\noindent
\textbf{Compared to LLM-assisted Static Auditing.} 
\chong{\gptscan discovers only 4 \high and 2 \medium vulnerabilities because it matches contracts against a fixed set of predefined scenarios, missing logic flaws outside these templates.
\llmaudit attains the highest recall among baselines (12 \high and 12 \medium) through an LLM-based workflow that iteratively reasons about and self-validates candidate vulnerabilities, yet it relies solely on the LLM's parametric knowledge, elicited via chain-of-thought and a few generic in-context examples.
\tool surpasses them by grounding detection in a large knowledge graph that maps DeFi semantics and linked vulnerability patterns onto the target project. Both baselines also incur many false positives (30 and 43): since neither executes the contracts, they cannot tell whether a reported finding is actually reachable, and their purely textual reasoning is prone to hallucination. In contrast, the agentic framework of \tool validates every finding with an executable PoC, driving false positives to zero.}

\noindent
\textbf{Compared to Coding Agent Harness.} 
\chong{
\codexevm detects 4 \high and 7 \medium vulnerabilities while achieving the highest precision among all baselines, with only six false positives. As an industry-grade coding agent, it autonomously explores the codebase using diverse tools and validates candidate vulnerabilities by executing PoCs through a CLI, contributing to its balanced precision and recall. Nevertheless, \codexevm remains substantially behind \tool in both precision and recall.
For precision, \codexevm lacks explicit vulnerability oracles, leaving the decision of whether an observed behavior constitutes a vulnerability entirely to the agent.
In contrast, \tool encodes the \textit{Pre-/Post-Vuln} invariants as executable \textit{require} oracles and validates them against concrete execution traces. For recall, the gap stems from both knowledge and workflow. On the knowledge side, augmenting \codexevm with our knowledge graph (\toolcodex) increases the number of detected \high vulnerabilities from 4 to 10, indicating that knowledge-guided coverage is a key contributing factor. On the workflow side, \tool systematically enumerates and verifies all candidate semantic--vulnerability pairs, ensuring broad coverage. In contrast, \codexevm autonomously controls its exploration process and cannot guarantee that all relevant vulnerability patterns are examined.
}

\noindent
\textbf{False Negatives of \tool.} 
\chong{Although \tool achieves strong detection performance, it still misses 6 \medium vulnerabilities. Our analysis of the knowledge graph and execution traces shows that these missed cases are typically tied to project-specific implementation details~\cite{web3bugs} or uncommon DeFi semantics, making them difficult to capture using abstract knowledge distilled from historical audits.}

\subsection{RQ2: Ablation Study}\label{sec:kg-cov}
\chong{In this section, we evaluate the contributions of the knowledge graph and the agentic workflow to the overall performance of \tool, and manually assess the quality of the knowledge graph constructed by \tool.}

\noindent\chong{\textbf{Contributions of Key Components.} Table~\ref{tab:rq1-bug-finding} reports the results of the two ablation variants, \toolnkg and \toolcodex. Removing the knowledge graph from the agentic workflow (\toolnkg) reduces the number of detected vulnerabilities by 81\% while maintaining high precision. This result highlights the critical role of the knowledge graph in improving detection recall and, conversely, demonstrates that the agentic workflow alone is effective at preventing false positives. The results of \toolcodex further illustrate the complementary contributions of the two components. Despite leveraging the same knowledge graph, \toolcodex achieves substantially lower precision and recall than \tool, indicating that \tool's agentic workflow is better suited to end-to-end smart contract auditing than a powerful general-purpose coding agent harness. At the same time, \toolcodex outperforms \codexevm, confirming that the knowledge graph provides substantial benefits regardless of the underlying agentic framework.}


\noindent \textbf{Quality of Knowledge Graph.}
\chong{
We randomly sample subsets of knowledge artifacts, including DeFi semantics, vulnerability patterns, and semantic--pattern links, from the constructed knowledge graph, ensuring a 95\% confidence level with a 5\% margin of error~\cite{cochran1977}. Two auditors with at least two years of Solidity auditing experience independently label each artifact as valid or invalid, and disagreements are resolved by a third expert auditor to produce the final adjudicated labels. We evaluate the quality of the knowledge graph using the precision of each artifact type and the inter-annotator agreement measured by Cohen's $\kappa$~\cite{cohen1960,landis1977}. All three artifact types achieve high precision with substantial agreement: DeFi semantics attain 87.6\% precision ($\kappa=0.82$), vulnerability patterns 98.3\% ($\kappa=0.80$), and semantic--pattern links 86.6\% ($\kappa=0.76$). These results demonstrate that \tool constructs a high-quality knowledge graph.
}

\subsection{RQ3: Unknown Vulnerability Discovery}\label{sec:rq3}
\chong{To evaluate the effectiveness of \tool in discovering previously unknown vulnerabilities, we apply it to seven newly developed real-world projects before their online deployment\footnote{Some projects were already deployed after our security audit. To preserve anonymity during peer review, we omit their identities and will disclose them upon acceptance.}.}

\noindent\textbf{Results.} \tool discovers 9 \high-, 36 \medium-, and 72 \low-severity vulnerabilities, all of which were confirmed and fixed by the corresponding development teams. The affected projects collectively secure over \$51.6M in on-chain total value locked (TVL)~\cite{defillama}. A detailed breakdown of all discovered vulnerabilities is provided in the artifact~\cite{figshare}.

\noindent
\textbf{Case Study.}
Figure~\ref{fig:rw-code} illustrates a previously unknown \high vulnerability in a protocol that authorizes privileged actions through off-chain signatures.
Each signed message carries a \texttt{deadline} that the contract checks against the current time, thereby ensuring the authorization expires after a fixed window.
For this to be effective, the \texttt{deadline} must itself be part of the signed data, so that the signer fixes the expiry and no relayer can change it later.
The \textit{Knowledge Mapper} of \tool matches this logic to a vulnerability pattern distilled from past audit contests that share no code with this project: \textit{a signed authorization whose expiry deadline is not actually bound by the signature}.
Guided by this knowledge, \tool finds that the \texttt{deadline} is a caller-supplied field that the signature does not cover: changing it does not invalidate the signature, so anyone relaying the signature can set the \texttt{deadline} to an arbitrary future time while it still verifies, as shown in Figure~\ref{fig:rw-code}.
The contract's expiry check is thus defeated, and a leaked or forwarded signature never truly expires.
\tool then synthesizes a proof-of-concept that replays the signature with an extended \texttt{deadline}, observes it still succeeds, and reports the finding, which the client confirmed as a \high vulnerability.

\begin{figure}[t]
    \centering
\begin{minted}[escapeinside=||,xleftmargin=0em,fontsize=\scriptsize]{solidity}
function execute(uint256 action, Sig calldata sig) {
    // deadline is checked against the clock...
    require(sig.deadline >= block.timestamp);
    // ...but it is left out of the signed hash, which
    // should be keccak256(abi.encode(action, sig.deadline))
    bytes32 h = keccak256(abi.encode(action));
    require(signer == ecrecover(h, sig.v, sig.r, sig.s));
}
\end{minted}
    \caption{The simplified vulnerable code of a previously unknown \high vulnerability \tool finds.}
    \label{fig:rw-code}
\end{figure}

\begin{figure}[t]
    \centering
    \includegraphics[width=0.8\linewidth]{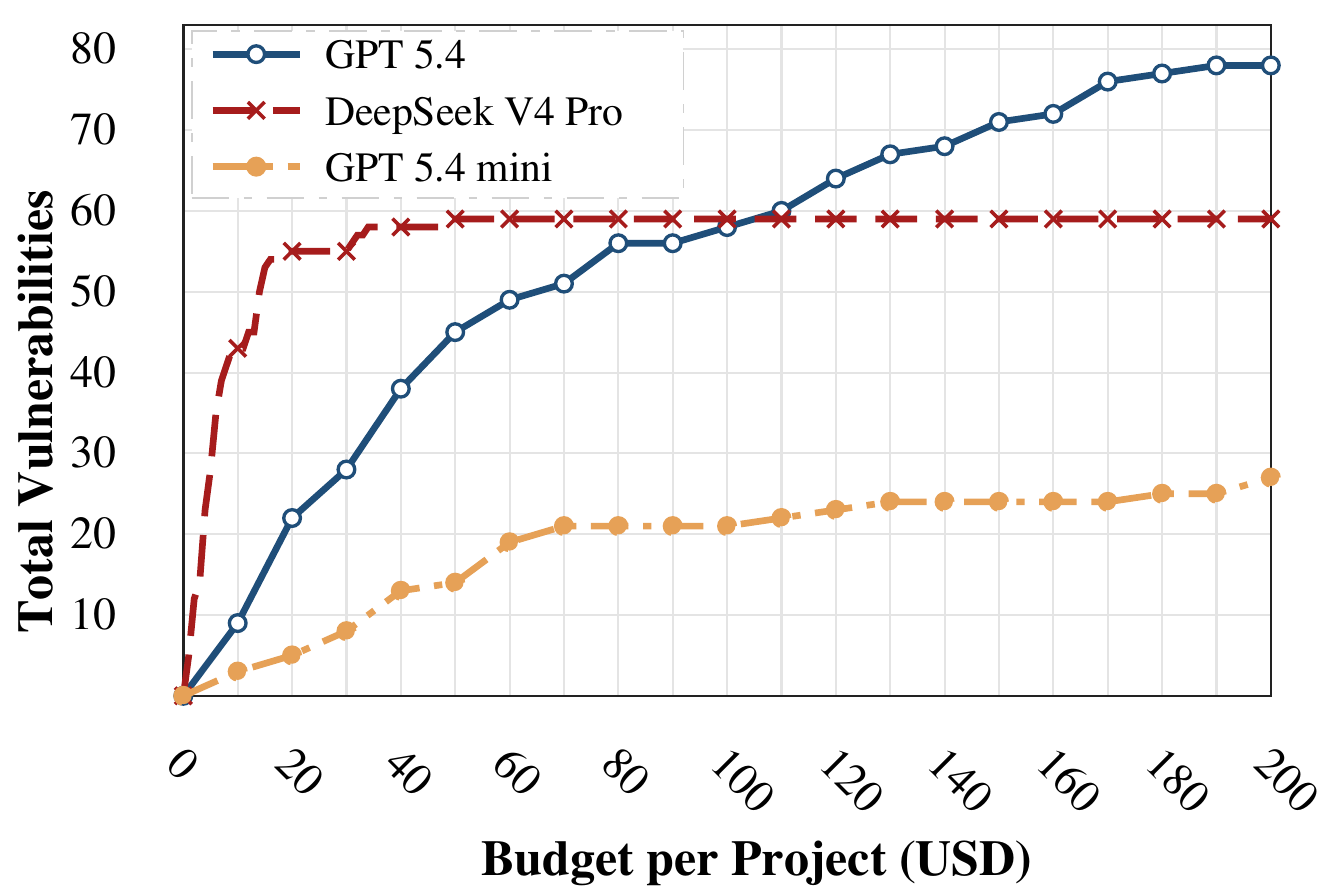}
    \caption{Cumulative vulnerabilities discovered on \eval under different per-project token budgets.}
    \label{fig:rq4-budget-bug-curve}
\end{figure}

\subsection{RQ4: Model Sensitivity Analysis}\label{sec:eval-model-sensitivity}
\chong{We evaluate \tool with two additional foundation models, \gptfivemini and \dpskpro~\cite{deepseekv4}, to study its sensitivity to the choice of foundation model in terms of both vulnerability detection effectiveness and token cost efficiency. 
}

\noindent
\textbf{Results.} 
\chong{
Figure~\ref{fig:rq4-budget-bug-curve} shows the cumulative number of vulnerabilities discovered on \eval under different token budgets. A detailed breakdown is provided in the artifact~\cite{figshare}. Among the three models, \gptfivefour achieves the highest vulnerability detection effectiveness when allocated a sufficient token budget, whereas \dpskpro offers the best token efficiency while maintaining relatively high vulnerability coverage. By inspecting the execution transcripts, we find that the primary reason for the lower detection ceilings of \dpskpro and \gptfivemini is that they often cause the \textit{Knowledge Mapper} to fail to concretize abstract security knowledge into project-specific auditing specifications. Overall, if maximizing vulnerability coverage is the primary objective, \gptfivefour is the preferred choice. If a better balance between vulnerability coverage and cost efficiency is desired, \dpskpro provides a more favorable trade-off.
}
\section{Discussion}
We highlight several actionable directions for improving the combination of knowledge extraction and LLM-based agents in smart contract auditing.

\noindent
\textbf{Continuous Knowledge Evolution.} An autonomous agent could periodically fetch the latest audit reports or related documentation \chong{from existing or new projects} to incrementally update the knowledge graph, keeping it up-to-date with emerging vulnerabilities and techniques.

\noindent
\textbf{Integration of Procedural Knowledge.} The current knowledge graph primarily captures conceptual and factual knowledge. Future work could incorporate procedural knowledge by distilling auditing skills or strategies from reports to guide specification generation and PoC synthesis more effectively.

\noindent
\textbf{Cost-Efficient Model Adaptation.} Our RQ4 results suggest that different LLMs excel at different aspects, revealing a promising direction for improving cost efficiency. Specifically, more capable but expensive models can be reserved for critical stages, such as auditing specification generation, while more economical models can be employed for routine code-related tasks, including code understanding and PoC generation.

\noindent
\textbf{Human-in-the-Loop Refinement.} While \tool demonstrates strong automated capabilities, lightweight human oversight could help resolve corner cases where LLM hallucinations or ambiguous specifications produce false positives or missed vulnerabilities. This hybrid approach could enhance reliability without substantially increasing cost.

\section{Threats to Validity}
\textbf{Internal Threats.} Internal validity threats primarily arise from the randomness of LLM outputs and the subjectivity of human evaluation. For human evaluation, we involve multiple annotators, provide detailed guidelines, and measure inter-annotator agreement to ensure consistency and reduce bias.

\noindent
\textbf{External Threats.} External validity threats mainly stem from the limited exploration of model selection in our study. Our experiments are conducted using \gptfivefour, \gptfivemini, and \dpskpro, so the results may not generalize to other LLMs. Another threat comes from our evaluation dataset, which is collected from Code4rena, and the findings may not generalize to auditing results from other platforms.
\section{Related Work}

\noindent
\textbf{LLM smart contract auditing.}
Recent work on LLM-based smart contract auditing has quickly moved from direct prompting over contract code to workflows grounded in richer audit evidence.
Much of the early work\cite{auditgpt,gptscan} is purely the static approach and shows that LLMs can reason about ERC rules and business-logic vulnerabilities directly from contract code, while several later works~\cite{logicscan,ckgllm,etrace,smartinv,propertygpt} further ground the model in mined specifications, knowledge graphs, traces, or retrieved human-written properties instead of relying on raw source alone.
More recent systems bring stronger feedback into the loop: LLM-SmartAudit~\cite{llmaudit} organizes the audit process as multi-agent collaboration, PromFuzz~\cite{promfuzz} uses LLM-generated bug-oriented analyses to guide functional bug detection, and SmartPoC~\cite{smartpoc} turns audit reports into executable PoCs.

\noindent
\textbf{Retrieval-augmented reasoning for code and security tasks.}
Retrieval-augmented generation~\cite{rag} has become a common way to compensate for the limited context and unstable recall of LLMs.
In code intelligence, this trend appears in repository-level retrieval systems such as RepoCoder~\cite{repocoder} and Repoformer~\cite{repoformer}, while GraphRAG~\cite{graphrag} suggests that graph-structured retrieval can support broader reasoning over large corpora.
The same idea is now widely used in security tasks, including Synapse~\cite{thought}, PropertyGPT~\cite{propertygpt}, Vul-RAG~\cite{vulrag}, VulInstruct~\cite{vulinstruct}, and CodeGuarder~\cite{codeguarder}.
\tool at a high level also fetches context from a knowledge graph.
However, the key difference is that the content retrieved by \tool is always semantic-aware instead of text similarity.

\noindent
\textbf{Fuzzing and proof-of-concept generation.}
Fuzzing and PoC generation focus on turning vulnerability hypotheses into executable evidence.
Recent LLM-based fuzzing work has started to generalize this process across domains~\cite{fuzz4all,hgfuzzer,elfuzz}.
In smart contracts, sFuzz~\cite{sfuzz}, Smartian~\cite{smartian}, ConFuzzius~\cite{confuzzius}, ItyFuzz~\cite{ityfuzz}, VERITE~\cite{verite}, and Belobog~\cite{belobog} show the importance of stateful exploration and domain-specific constraints.
PoC-oriented work further synthesizes exploits or validation scripts from known weaknesses or reports~\cite{goodnewsaeg,prompttopwn,smartpoc,webpocllm}.
\tool focuses on extracting expert audit knowledge and applying it to new smart contract projects, while concrete PoC execution is used to validate candidate findings.
\section{Conclusion}

We present \tool, a knowledge-driven, agentic framework for smart contract vulnerability detection. By constructing an auditing knowledge graph that links fine-grained DeFi semantics with recurring vulnerability patterns, and leveraging it through an iterative multi-agent auditing loop, \tool enables systematic, automated auditing and outperforms all baselines on our evaluation dataset.


\bibliographystyle{IEEEtran}
\bibliography{acmart}
\balance
\end{document}